\documentclass{desyproc}
\pdfoutput=1
\begin{document}
\title{Preliminary Results of the CASCADE Hidden Sector Photon Search}

\author{{\slshape  Nathan Woollett$^{1,2}$, Ian Bailey$^{1,2}$, Graeme Burt$^{1,2}$, Swapan Chattopadhyay$^{6,7}$, John Dainton$^{2,5}$, Amos Dexter$^{1,2}$, Phillippe Goudket$^{2,3}$, Michael Jenkins$^{1,2}$,  Matti Kalliokoski$^4$, Andrew Moss$^{2,3}$, Shrikant Pattalwar$^{2,3}$, Trina Thakker$^{2,3}$, Peter Williams$^{2,3}$}
\\[1ex]
$^1$Lancaster University, Lancaster, United Kingdom\\
$^2$The Cockcroft Institute of Accelerator Science and Technology, Warrington, United Kingdom\\
$^3$STFC ASTEC, Sci-Tech Daresbury, Warrington, United Kingdom\\
$^4$CERN, Geneva, Switzerland\\
$^5$University of Liverpool, Liverpool, United Kingdom\\
$^6$Northern Illinois University, Illinois, United States of America\\
$^7$Fermilab, Illinois, United States of America\\
}

\contribID{familyname\_firstname}

\confID{11832}  
\desyproc{DESY-PROC-2015-02}
\acronym{Patras 2015} 
\doi  

\maketitle

\begin{abstract}
Light shining through a wall experiments can be used to make measurements of photon-WISP couplings. The first stage of the CASCADE experiment at the Cockcroft Institute of Accelerator Science and Technology is intended to be a proof-of-principle experiment utilising standard microwave technologies to make a modular, cryogenic HSP detector to take advantage of future high-power superconducting cavity tests. In these proceedings we will be presenting the preliminary results of the CASCADE LSW experiment showing a peak expected exclusion of $1.10\times10^{-8}$ in the mass range from 1.96\,$\mu$eV to 5.38\,$\mu$eV, exceeding current limits.
\end{abstract}

\section{Introduction}

CASCADE (CAvity Search for Coupling of A Dark sEctor) is an experiment which utilises microwave cavities and amplifiers to search for energy transmission between the cavities beyond that which would be expected through Standard Model processes. This approach is sensitive to hidden sector photons (HSPs) through their kinetic mixing with the photon as described by the Lagrangian,

\begin{equation}
\mathcal{L}=-\dfrac{1}{4}F^{\mu\nu}F_{\mu\nu}-\dfrac{1}{4}B^{\mu\nu}B_{\mu\nu}-\dfrac{1}{2}\chi F^{\mu\nu}B_{\mu\nu}-\dfrac{1}{2}m^{2}_{\gamma '}B_\mu B^\mu,
\end{equation}

where $\chi$ is the coupling factor, $F$ is the standard model electromagnetic field and $B$ is the HSP field.

The method being used is known as a \emph{light shining through a wall}, LSW, experiment. In this design a cavity is powered from an external RF source and a second cavity is shielded from the powered cavity and used as a detector for transmission between the cavities. The cavities are nominally identical and operate at the same resonant frequency. This enables us to look for an excess of power in the detector and if the corresponding frequency matches that of our source, we can conclude that the  excess may originate from photon-HSP oscillations. A more detailed description of the technique is given in \cite{Jaeckel:2007ch}

\section{Measurement Set-up}

\begin{figure}

\centering
\hspace*{\fill}
\includegraphics[width=0.45\textwidth]{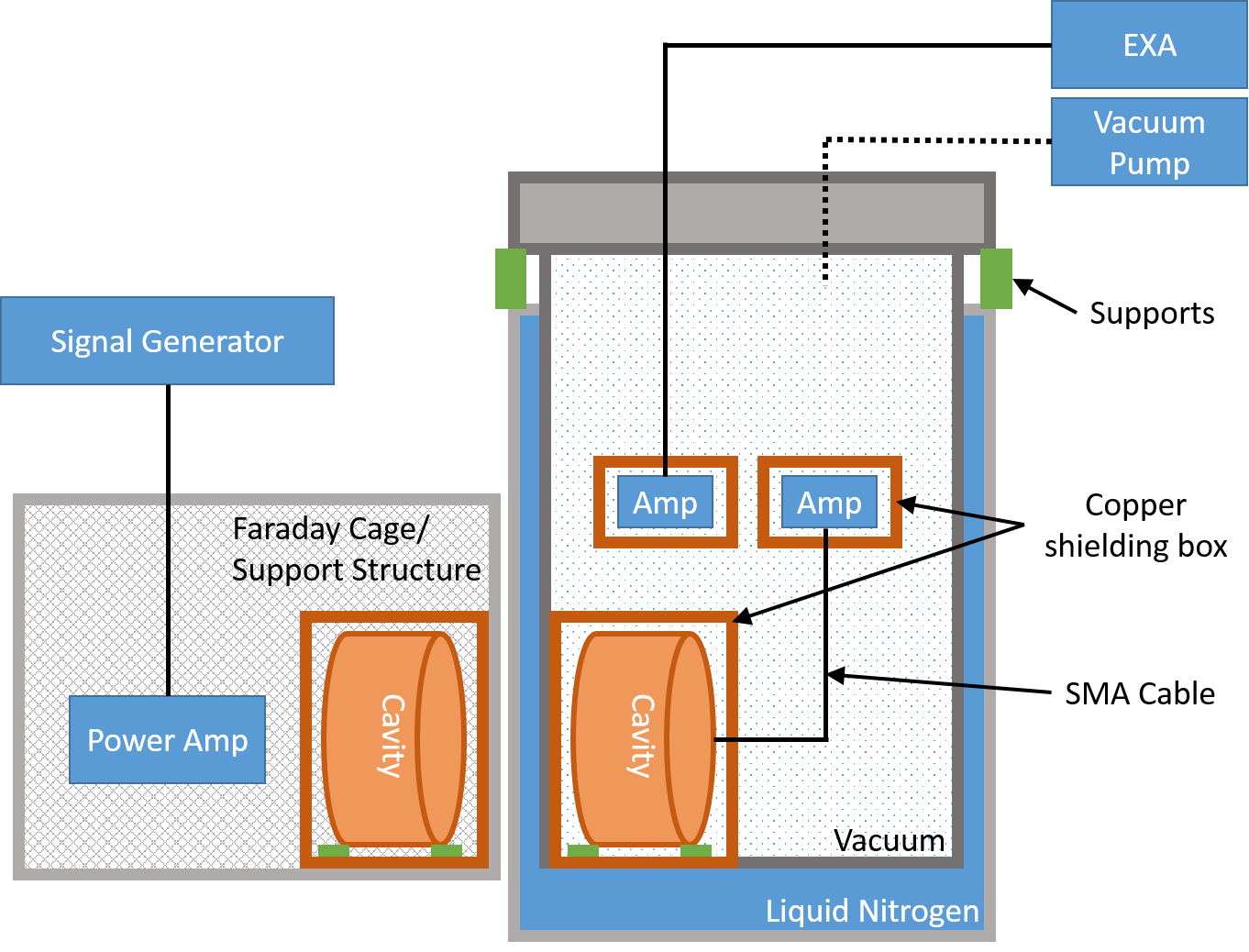}
\hfill
\includegraphics[width=0.45\textwidth]{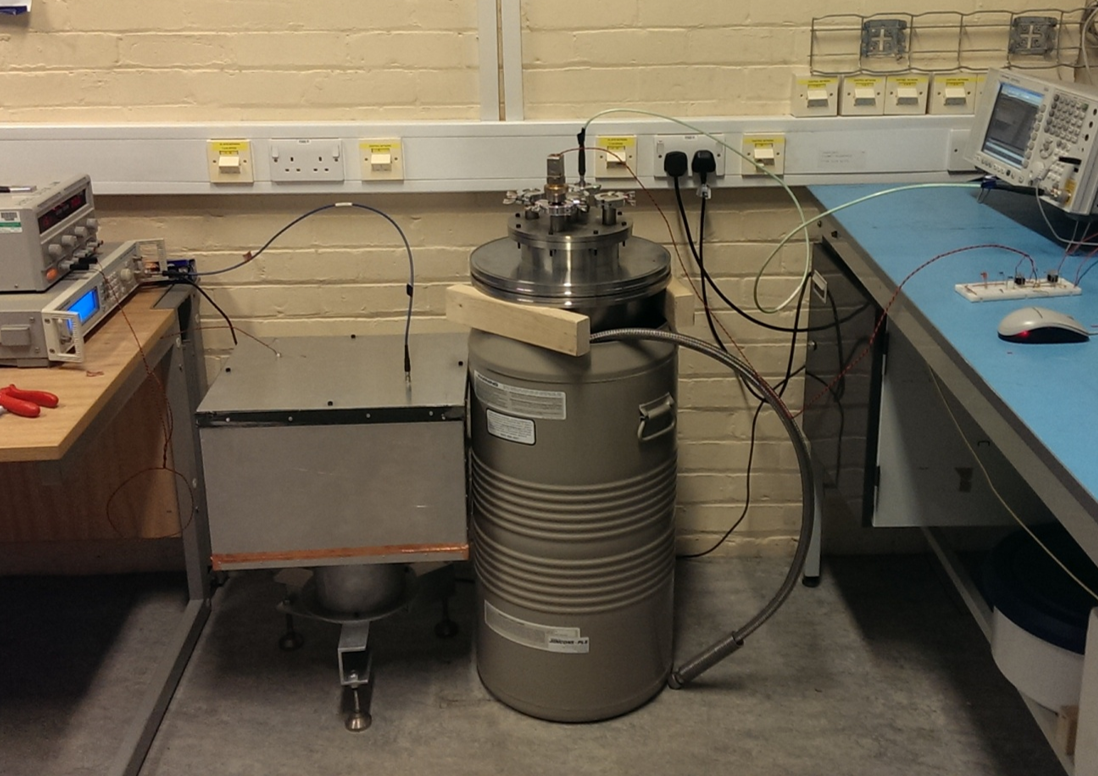}
\hspace*{\fill}
\caption{\emph{Left:} A schematic of the CASCADE experiment. On the left hand side is the emitter chain and on the right is the detector chain. Spatial positioning is representative of the final set-up but not to scale. \emph{Right:} A photograph of the final CASCADE set-up that was used to make a measurement. The photograph is taken from the same perspective as the schematic for easy comparison.}
\label{fig:Schematic}
\end{figure}

The CASCADE experiment employs two cavities and uses amplifiers to maximise the observable power. To minimise any transmission between the emitter and the detection system, care is taken to separate the two systems. A schematic of the set-up is shown in Fig.\ref{fig:Schematic}.

The emitter system consists of a signal generator, a power amplifier and a copper cavity. There are two layers of shielding within this chain, a copper box containing the cavity and an aluminium box around the cavity and power amplifier. By having a shielding box around the amplifier the RF power in the cables between the signal generator and the shielding box is kept at -4\,dBm rather than the 28\,dBm that was provided to the emitter cavity. The signal generator shares a common 10\,MHz reference signal with the detector chain to ensure frequency lock between the two systems. 

The detector chain consists of an Agilent Technologies EXA Signal Analyser, two Miteq ASF3 Cryogrenic Amplifiers and a copper cavity. There are two layers of shielding around the cavity with a copper box around the cavity and a stainless steel vacuum box forming a second box which also contains the amplifiers. It was found that if the amplifiers were themselves unshielded anomalous signals would be produced by cross talk between the amplifiers so individual copper boxes are used to shield them from one another.

The cavities were designed to operate in the first transverse magnetic mode (TM010) at 1.3\,GHz. The quality(Q) factor was estimated to be up to 22000 at room temperature using simulations in CST Microwave Studio, however, as the copper used is not oxygen free, the Q factor is limited to approximately $10500$.

The amplifiers were chosen for their low noise characteristics with a noise figure of 0.6\,dB at room temperature and as low as 0.2\,dB at cryogenic temperatures\cite{miteq}. The frequency region of interest is of the order of 100\,MHz, limited by the cavity tuning, and the observation window is only 80\,Hz, required to achieve the desired frequency resolution. The change in amplification with frequency for the amplifiers was found to be negligible over this range. Another important feature is their gain of 38\,dB, this was tested by recording the observed power as a function of input power. A small increase in amplification was observed with reducing input power but, as the exact cause of the increase was unknown, the minimum observed value of 38\,dB per amplifier was used in the calculating limits.  By having two amplifiers in series we can amplify the thermal noise floor which is estimated to be -230\,dBm above the internal noise of the signal analyser which is approximately -160\,dBm.

\section{Results}

\begin{figure}
\centering
\includegraphics[width=0.8\textwidth]{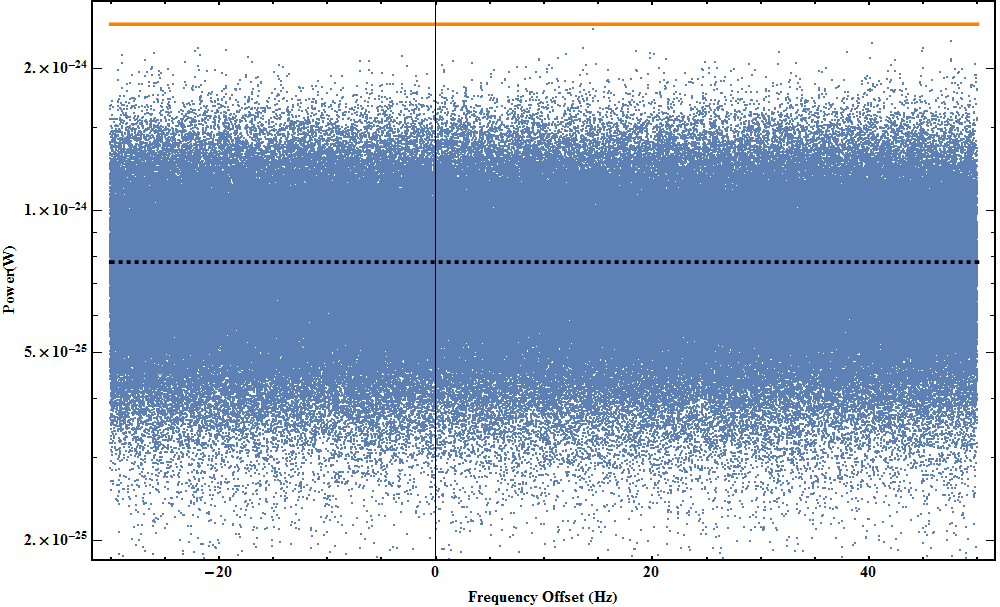}
\caption{The signal recorded during a CASCADE run. The dashed line indicates the mean noise power and the solid orange line indicates 5 standard deviations. It can be seen that there are no points in excess of this level which indicates a null observation.}
\label{fig:Noise}
\end{figure}

To make an exclusion we need to estimate the smallest signal we would be sensitive to. Since we are using a light shining through a wall experiment the signal frequency is known to be 1.293539940\,GHz meaning we can use the side bands in the measurement to estimate the noise level in the signal region. The recorded data was over-sampled giving 524288 points each sensitive to a bandwidth of 0.5\,mHz. A 5 sigma confidence level was used as this gives a less than 0.3\% probability of having an excess within 1\,Hz of the signal frequency assuming a flat noise distribution. No excess of power was recorded as can be seen in Fig. \ref{fig:Noise} where the 5 sigma power level is indicated by the solid orange line. 

\begin{figure}
\centering
\includegraphics[width=0.8\textwidth]{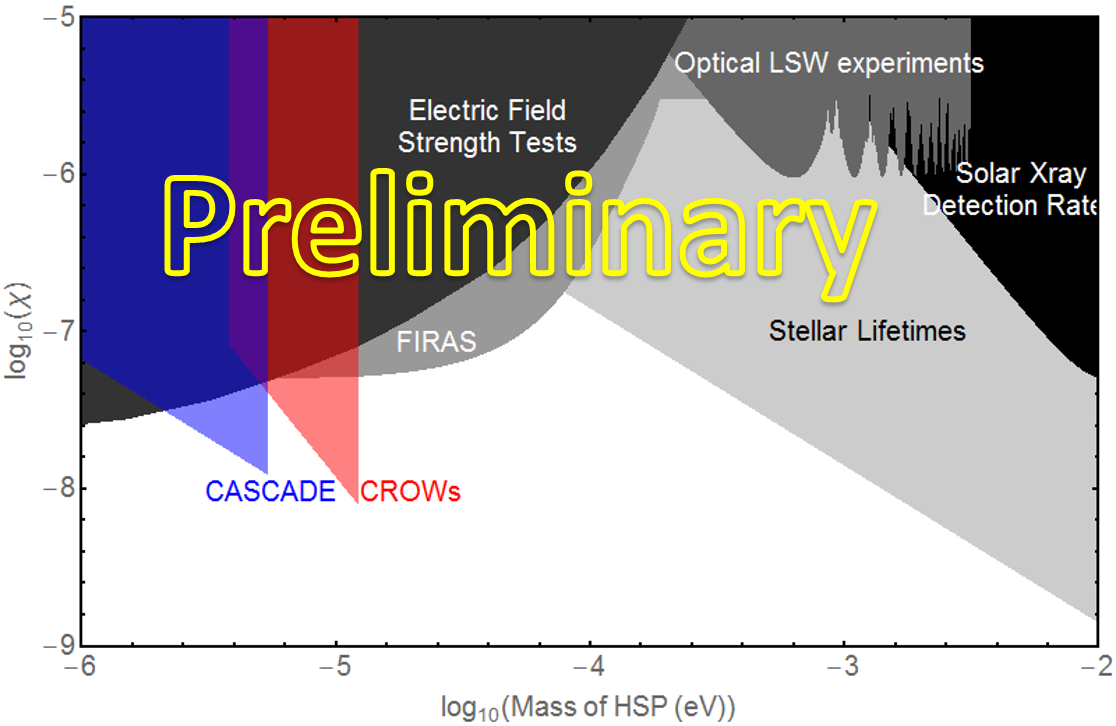}
\caption{The expected exclusion of the CASCADE experiment based on the far field approximation of cavities coupled through the longitudinal polarisation of the HSP. The CROWs experiment has been highlighted for comparison as it employs the same experimental technique.}
\label{fig:Results}
\end{figure}

The cavities used in CASCADE are aligned coaxially. This enables us to take advantage of the longitudinal polarisation mode of the HSP. For the longitudinal mode the sensitivity is proportional to $(m_{\text{HSP}}/\omega)^{2}$, rather than $(m_{\text{HSP}}/\omega)^{4}$ as for the transverse mode  where $m_{\text{HSP}}$ is the mass energy of the HSP and $\omega$ is the energy of the photons within the source\cite{Graham:2014sha}. This enables the search to cover lower masses leading to the mass range where the new parameter space is covered to increase from 1.1\,$\mu$eV to 3.4\,$\mu$eV. The preliminary exclusion based on the far field approximation of HSP coupling is shown in Fig. \ref{fig:Results} presenting a peak exclusion down to a mixing factor, $\chi$, of $1.10\times10^{-8}$ and the strongest exclusion from 1.96\,$\mu$eV to 5.38\,$\mu$eV. However the cavity separation was on the order of the cavity height which is closer than appropriate for the far-field approximation so the final result will be based on a full near-field calculation.

\section{Acknowledgments}

This research was funded in part through the STFC Cockcroft Institute Core grant no. ST/G008248/1.




\begin{footnotesize}

\end{footnotesize}


\end{document}